\documentclass[conference]{IEEEtran}
\IEEEoverridecommandlockouts
\usepackage{cite}
\usepackage{amsmath,amssymb,amsfonts}
\usepackage{algorithmic}
\usepackage{graphicx}
\usepackage{textcomp}
\usepackage{xcolor}
\def\BibTeX{{\rm B\kern-.05em{\sc i\kern-.025em b}\kern-.08em
    T\kern-.1667em\lower.7ex\hbox{E}\kern-.125emX}}
\begin{document}

\title{A Cognitive Approach to Improving Binary Reverse Engineering with Immersive Virtual Reality\\
}

\author{\IEEEauthorblockN{1\textsuperscript{st} Dennis G. Brown}
\IEEEauthorblockA{\textit{Comp. Sci. and Software Eng.} \\
\textit{Auburn University}\\
Auburn, AL, USA \\
dgb0028@auburn.edu}
\and
\IEEEauthorblockN{2\textsuperscript{nd} Julian Bauer}
\IEEEauthorblockA{\textit{Comp. Sci. and Software Eng.} \\
\textit{Auburn University}\\
Auburn, AL, USA \\
jcb0209@auburn.edu}
\and
\IEEEauthorblockN{3\textsuperscript{rd} Luke Wittbrodt}
\IEEEauthorblockA{\textit{Comp. Sci. and Software Eng.} \\
\textit{Auburn University}\\
Auburn, AL, USA \\
ljw0024@auburn.edu}
\and
\IEEEauthorblockN{4\textsuperscript{th} Samuel Mulder}
\IEEEauthorblockA{\textit{Comp. Sci. and Software Eng.} \\
\textit{Auburn University}\\
Auburn, AL, USA \\
szm0211@auburn.edu}
}

\maketitle

\begin{abstract}
Through its affordances, immersive virtual reality (VR) offers a means to apply embodied and external cognition from the physical realm to solving analytical problems that are typically only conceptual. We present an example of executing a structured analysis following the tenets of cognitive systems engineering to derive immersive affordances applicable to a difficult analytical problem, in our case, reverse engineering (RE) binary programs. We conducted a basic cognitive task analysis of the problem to reveal features of its cognitive model and their associated fundamental cognitive phenomena, and then we mapped those concepts to immersive affordances associated with those concepts. We implemented a subset of those affordances in a VR system facilitating discovery of features of a binary program. Feedback from RE practitioners drove the initial development of the system and we are preparing for a formal effectiveness study to inform the direction of future research. 
\end{abstract}


\section{Introduction}
\label{section:introduction}

Binary reverse engineering (RE) --- understanding the capabilities of software distributed as binary code --- is critical to tasks such as securing networks and maintaining legacy software. For example, an organization may find suspected malware running on their network and will need to analyze the executable file to catalog its capabilities and start a remediation process. Many existing analysis tools, such as disassemblers, decompilers, profilers, and debuggers, offer insight into various aspects of binary programs; in combination, they provide significant, if disparate, insights into program behavior at a low semantic layer. However, binary RE is difficult~\cite{Meng2016}. Binary programs are created from source code in a lossy multi-step process, and reversing these steps introduces significant uncertainty at each step. This uncertainty thwarts fully automating binary RE --- the process requires a human in the loop.

\begin{figure*}[htbp]
\centerline{\includegraphics[width=0.67\textwidth]{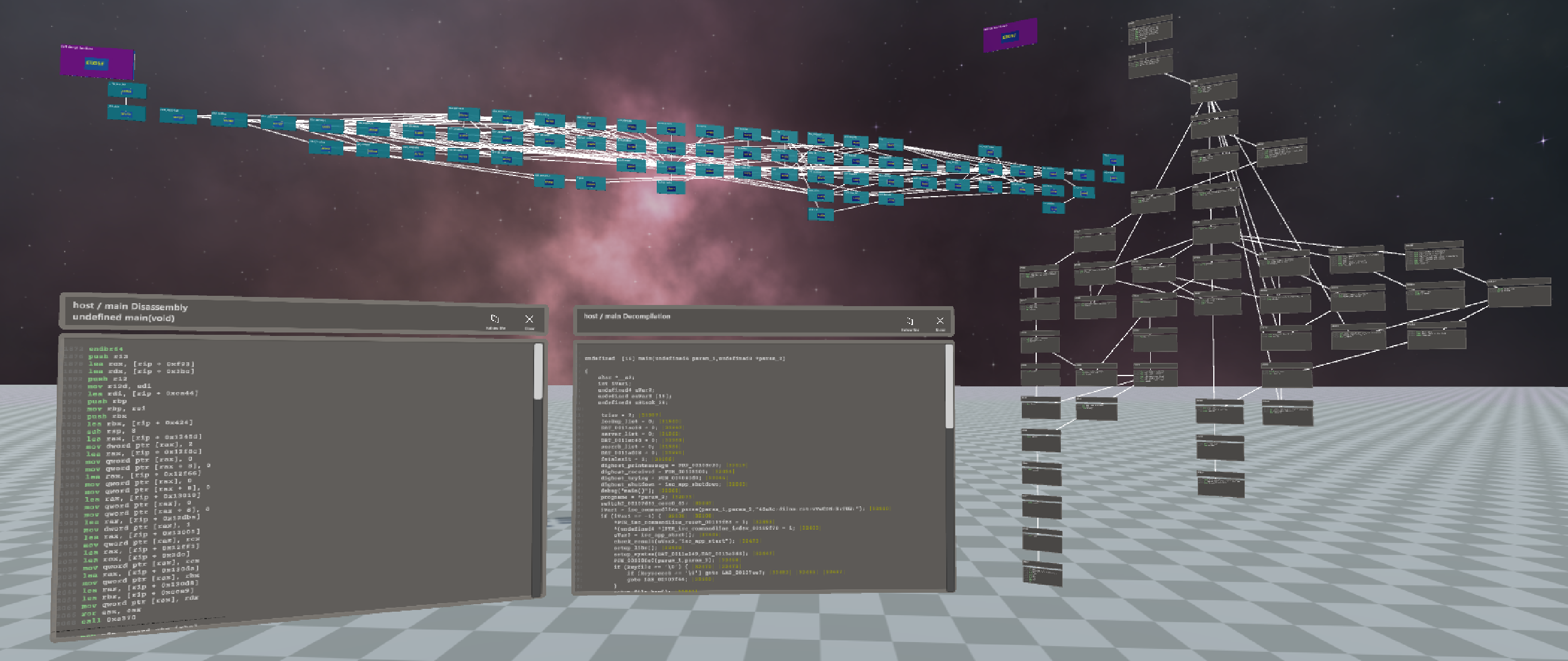}}
\caption{Exploring data recovered from a typical small binary program in our VR system: function call graph (upper left) and the disassembly, decompilation, and control flow graph for a selected function (left to right under function call graph).}
\label{fig:screenshot}
\end{figure*}

To improve this human-centric process, we are investigating how to reduce the cognitive load of binary RE. One compelling avenue is to tap into embodied and external cognition~\cite{Wilson2002} via virtual reality (VR). This paper summarizes our journey so far. We follow a cognitive systems engineering~\cite{Hollnagel2005} approach broken into two parts executed iteratively and in parallel. The first part is a basic cognitive task analysis of binary RE; we provide a high-level summary of that analysis in Section~\ref{sec:cta}. In the first iteration of the task analysis, we identified three primary themes of cognition where we see high value applying specific affordances in VR. The parallel second part is a design thinking approach to building a virtual reality testbed, which is covered in Section~\ref{sec:implementation}; see an example of a user session in Figure~\ref{fig:screenshot}. This is another iterative process where feedback from RE practitioners on the testbed drives feature development; Section~\ref{sec:feedback} discusses our first round of feedback. We are preparing to perform formal effectiveness evaluations with this testbed, as discussed in Section~\ref{sec:study}.

\section{Tracing Cognition to Affordances in VR}
\label{sec:cta}

To understand the binary RE process, we performed a basic form of cognitive task analysis~\cite{Clark2008} of the joint cognitive system of human and machine as part of our overall cognitive systems engineering approach described in the introduction. We performed a comprehensive literature survey~\cite{Brown2024arxiv} to provide deep preliminary knowledge of the binary RE task and cognitive models employed by practitioners. We also elicited task knowledge from lab members with experience up to the expert level. With this information, we performed conceptual clustering and extracted salient conceptual elements in three groups: cognitive models of binary RE, cognitive phenomena related to those elements, and affordances in VR that affect those cognitive phenomena. With these elements, we performed a synthesis and prioritization to determine the first set of affordances we would implement and test. In the interest of brevity, this paper only covers the elements directly relevant to our initial implementation; all elements are included in the above-referenced survey. We begin with discussing the elements of the first group, cognitive models of binary RE. 

\subsection{Cognitive Models of Binary RE}
\label{section:cogmodel}

Cognitive (or mental) models are developed by humans as internal representations to reason about happenings in the external world~\cite{Craik1943}. Researchers have developed several cognitive models of software RE and specifically binary RE, capturing the sensemaking process of determining the features and capabilities of code. Of the several published cognitive models of RE we reviewed, we found the following salient and common themes. 

In the most basic software RE, the practitioner makes sense of a program by starting with observation of the program. The fundamental cognitive model uses short term memory and existing knowledge of semantics (general computing concepts) and syntax (language-specific), in combination with that observation, to form a multi-level internal semantic representation of the program; this is the basis of program comprehension.~\cite{Shneiderman1979}

Later cognitive models of RE, more focused on binary RE, incorporate an iterative pattern of sensemaking or abductive reasoning, which involves repeatedly forming, testing, and updating hypotheses within the bounds of goals and plans~\cite{Bryant2012}~\cite{Nyre-Yu2022}. Practitioners form increasingly complete hypotheses~\cite{Brooks1983}~\cite{Weigand2012}~\cite{Dudenhofer2019}~\cite{Votipka2020} and test hypotheses through experimentation~\cite{Bryant2012}~\cite{Sisco2017}~\cite{Dudenhofer2019}~\cite{Votipka2020}. During this iteration, they update the framing of the problem based on observed results~\cite{Klein2007}~\cite{Bryant2012}~\cite{Dudenhofer2019}~\cite{Votipka2020}.

Researchers have observed other overarching characteristics of the binary RE process. Practitioners can become disoriented following recursions and execution paths~\cite{Zayour2000}. The process taxes working memory~\cite{Shneiderman1979}~\cite{Zayour2000}, and it often relies on using external memory aids~\cite{Detienne2001}~\cite{Storey2005} and on determining what to ignore~\cite{Mantovani2022}. Successful completion requires retrieving and generating declarative (factual) and procedural (patterns of interaction) knowledge~\cite{Bryant2012}. The process relies on referring to an overview of the binary executable~\cite{Votipka2020}. Many practitioners employ translation: determining how the code would be implemented in a higher-level language~\cite{Sisco2017}. Many also identify beacons, somewhat arbitrarily-defined items of interest, to guide their work, and beacons for binary RE are more diverse than for source-code-based PC~\cite{Brooks1983}~\cite{Dudenhofer2019}~\cite{Votipka2020}.

\subsection{Cognitive Theory Applied to Binary RE}
\label{section:cognition}

After discovering elements of cognitive models of RE, we researched the cognitive phenomena that impact the model elements. We found examples in the literature that fit into three general categories. 

\textbf{External Cognition}: Humans interact with external manifestations of knowledge in the world. One example is to use these external knowledge representations to reduce memory load, e.g., notes and reminders~\cite{Scaife1996}. Another example is using computational tools (e.g., calculators, or in our use case, various RE analysis tools) to make tasks easier~\cite{Preece2019}. A final example we consider is annotating, reordering, or restructuring external representations of knowledge~\cite{Preece2019}.

\textbf{Embodied Cognition and Memory}: Sensorimotor interaction with the environment impacts cognition as the brain processes inputs from the body and external environment. Specifically, building a \textit{memory palace} enhances memory by associating data with specific locations or objects in an environment, exploiting our natural reliance on navigational and spatial cognition~\cite{Ale2022}. Additionally, whole-body stimuli can expedite storage and retrieval of memory, for example, associating data with a particular motion or sensory input~\cite{Ale2022}.

\textbf{Cognitive Load Theory}: In this theory, humans have a limited processing ``bandwidth'' and that bandwidth can be optimized through balancing types of cognitive load. \textit{Intrinsic load} is inherent in the task. \textit{Germane load} connects current processing to long-term memory, developing schema. \textit{Extraneous load} is caused by the means of interacting and how a task is presented.~\cite{Sweller2019}. Several methods may reduce extraneous load, allowing for processing additional intrinsic or germane load: studying solved sample problems (worked example effect); presenting multiple information sources through different modalities (e.g., visual and aural) to allow the inputs in each modality to be processed simultaneously (modality effect); and removing redundancy of information presented in different modalities/sources to reduce the load of reconciling the underlying concepts across the inputs from those modalities~\cite{Hollender2010}.

\begin{figure*}
\centerline{\includegraphics[width=0.90\textwidth]{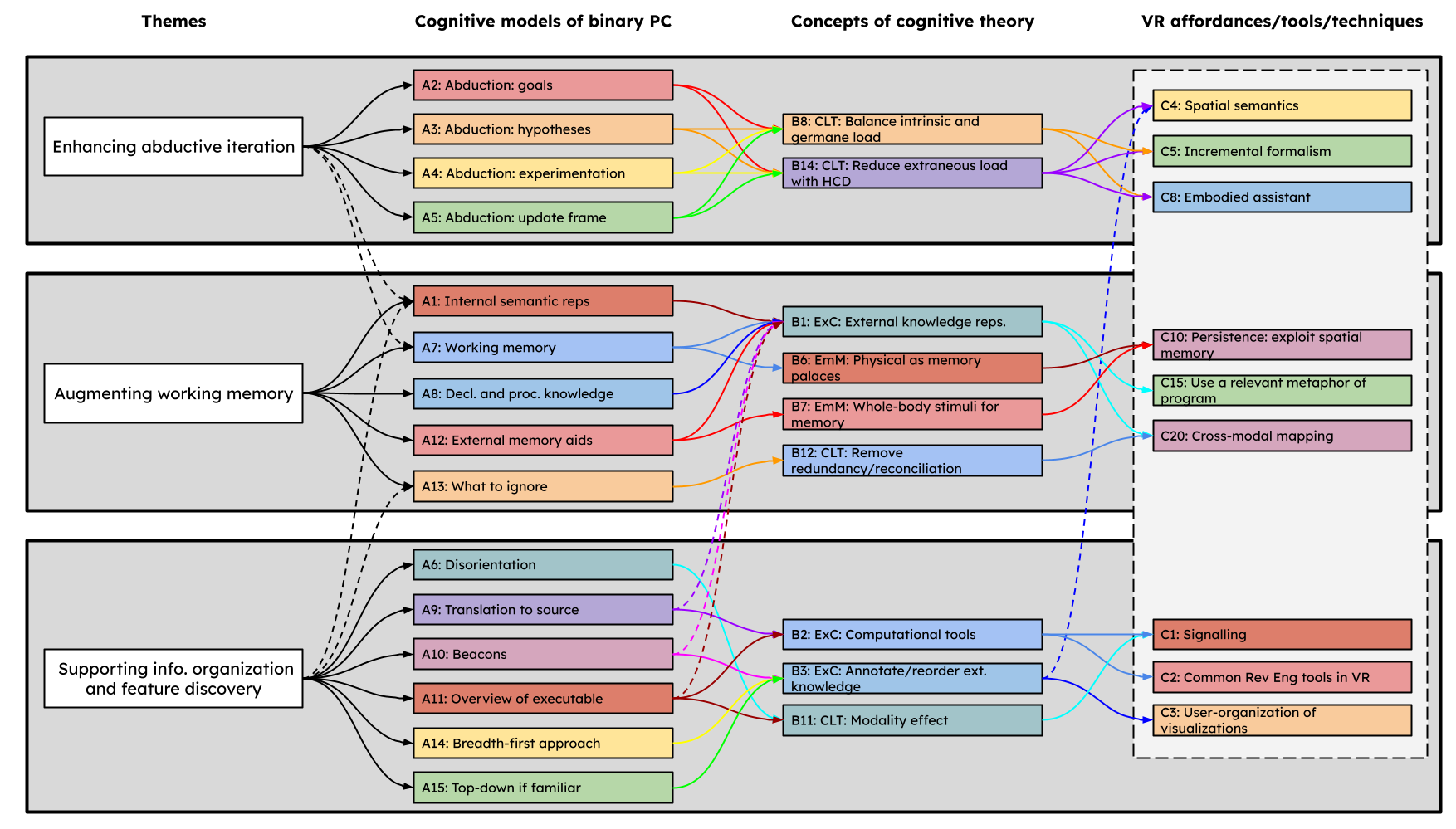}}
\caption{Primary themes for analysis with most closely-related elements; highlighted area indicates highest-priority affordances in VR.}
\label{fig:themes}
\end{figure*}

\subsection{Synthesis: Identifying Related Affordances in VR}
\label{sec:affordances}

After identifying the salient features of the RE sensemaking process (first group of elements) and their associated cognitive phenomena (second group of elements), we surveyed the affordances of immersive VR most closely associated with these phenomena (third group of elements), including prior work in applying VR to RE. The elements in this third group will be described below in the context of themes. 

We must clarify that in this paper, ``affordances in VR'' refers to the environmental features available in the VR experience. Some of these features are intrinsically 2D, such as code listing windows, but they can be created, sized, oriented, and placed arbitrarily in the 3D space by the user as they carry out a task. This approach builds a foundation for evaluating the effects of external and embodied cognition, and is the subject of active research in other use cases~\cite{Lisle2021}.  

With these three groups of elements, we formed a set of \emph{threads} (also called \emph{strands} in similar work). Each thread connects a cognitive model element from the first group to a cognitive phenomenon from the second group that demonstrates that cognitive model element, and then traces that phenomenon to an affordance in the third group that significantly impacts that phenomenon. Through an affinity mapping process, we grouped the threads into three cohesive themes representing targets of opportunity, as depicted in Figure~\ref{fig:themes}. We describe these themes along with their constituent elements from first, second, and third groups.

\textbf{Enhancing abductive iteration}: The iterative pattern of sensemaking using abductive reasoning is common to most cognitive models of RE and consists of four iterative elements: \emph{setting goals/following plans, forming hypotheses, experimenting to test hypotheses, and updating what is known}~\cite{Bryant2012}~\cite{Nyre-Yu2022}~\cite{Brooks1983}~\cite{Weigand2012}~\cite{Dudenhofer2019}~\cite{Votipka2020}~\cite{Sisco2017}~\cite{Klein2007}. 
Cognitive theory elements most associated with these cognitive model elements are focused on cognitive load in the iterative sensemaking loop through \emph{balancing the intrinsic and germane loads}~\cite{Sweller2019} and by \emph{reducing extraneous load as much as possible through human-centered design}~\cite{Helgesson2021}. 
Further, the affordances in VR associated with the cognitive theory elements include \emph{incremental formalism}~\cite{Andrews2010}, in which the visualization of the iterative process evolves its structure as progress is made; \emph{spatial semantics}~\cite{Andrews2010}, in which the spatial organization of the information provides semantic information; and the \emph{embodied assistant}~\cite{deMelo2020}, an avatar visually present in the environment, that would track and inform the expert's decision path. 

\textbf{Augmenting working memory}: Using working memory effectively is critical to binary RE. Characteristics from cognitive models of RE most closely related to this theme include \emph{multi-level internal semantic representation}~\cite{Shneiderman1979}, \emph{taxing working memory}~\cite{Shneiderman1979}~\cite{Zayour2000}, \emph{generation, storage, and retrieval of declarative and procedural knowledge}~\cite{Bryant2012}, \emph{using external memory aids}~\cite{Detienne2001}~\cite{Storey2005}, and \emph{determining what to ignore}~\cite{Mantovani2022}. We can trace those model elements to facets of cognitive theory: \emph{using external knowledge representations to reduce memory load}~\cite{Scaife1996}~\cite{Preece2019} by offloading some knowledge that otherwise would be held in working memory, \emph{employing memory palaces aided by physical objects or locations} and \emph{leveraging whole-body stimuli to expedite storage and retrieval of memory}~\cite{Ale2022}, and \emph{removing redundancy of information presented in different modalities}~\cite{Hollender2010} to streamline the intake of new information into working memory. 
The affordances in VR related to these cognitive theory elements include \emph{employing persistence to exploit spatial memory}~\cite{Andrews2010} to remember information, \emph{using physical metaphors}~\cite{Fittkau2015}~\cite{Oberhauser2017}~\cite{Capece2017}~\cite{Averbukh2019}~\cite{Romano2019}~\cite{Hoff2022} for the task's information and operations, and \emph{cross-modal mapping}~\cite{Moloney2018} to present information through multiple senses. 

\textbf{Supporting information organization and feature discovery}: The binary RE process involves finding, characterizing, and reasoning about significant features of the program. Cognitive model elements closely tied to this theme include \emph{disorientation following execution paths}~\cite{Zayour2000}, \emph{translating the binary back to source code}~\cite{Sisco2017}, \emph{using or marking beacons}~\cite{Brooks1983}~\cite{Dudenhofer2019}~\cite{Votipka2020}, \emph{referring to an overview of the binary program}~\cite{Votipka2020}, and \emph{approaching the task in a breadth-first and top-down manner, depending on the expert's familiarity}~\cite{Vessey1985}~\cite{Siegmund2014}. 
These model elements trace to elements of cognitive theory. \emph{Annotating and reordering/restructuring external representations of
knowledge}~\cite{Scaife1996}~\cite{Preece2019} addresses note-taking and organizing artifacts. \emph{Using computational tools to make tasks easier}~\cite{Scaife1996}~\cite{Preece2019}, in this use case, includes the use of reverse engineering analysis tools to discover key characteristics of the binary. The \emph{modality effect}~\cite{Hollender2010} can take advantage of multiple types of information about the binary program to provide signalling cues. 
Several affordances in VR apply to this theme. First, \emph{incorporate common reverse engineering tools} in a way that allows the expert to naturally exploit their findings in the immersive environment. Second, \emph{signalling}~\cite{Albus2021}~\cite{Mayer2005} directs users to what is most important, either through cues set by the user or generated automatically from analysis tools. Finally, \emph{user organization of visualizations}~\cite{Batch2020} is one way that users can capture relationships between program features or notations in how they are organized in the space. 

NOTE: There are three similar affordances in VR across the themes that may seem redundant, so we want to differentiate them. \textit{Spatial semantics} exploits spatial position and representation to help the practitioner \textit{understand}. \textit{Persistence} exploits the same to help the practitioner \textit{remember}. \textit{User organization} exploits the same to help the user \textit{express knowledge}. All three affordances are part of exploiting the immersive space to improve performance~\cite{Lisle2021}.

\begin{figure}[htbp]
\centerline{\includegraphics[width=0.5\textwidth]{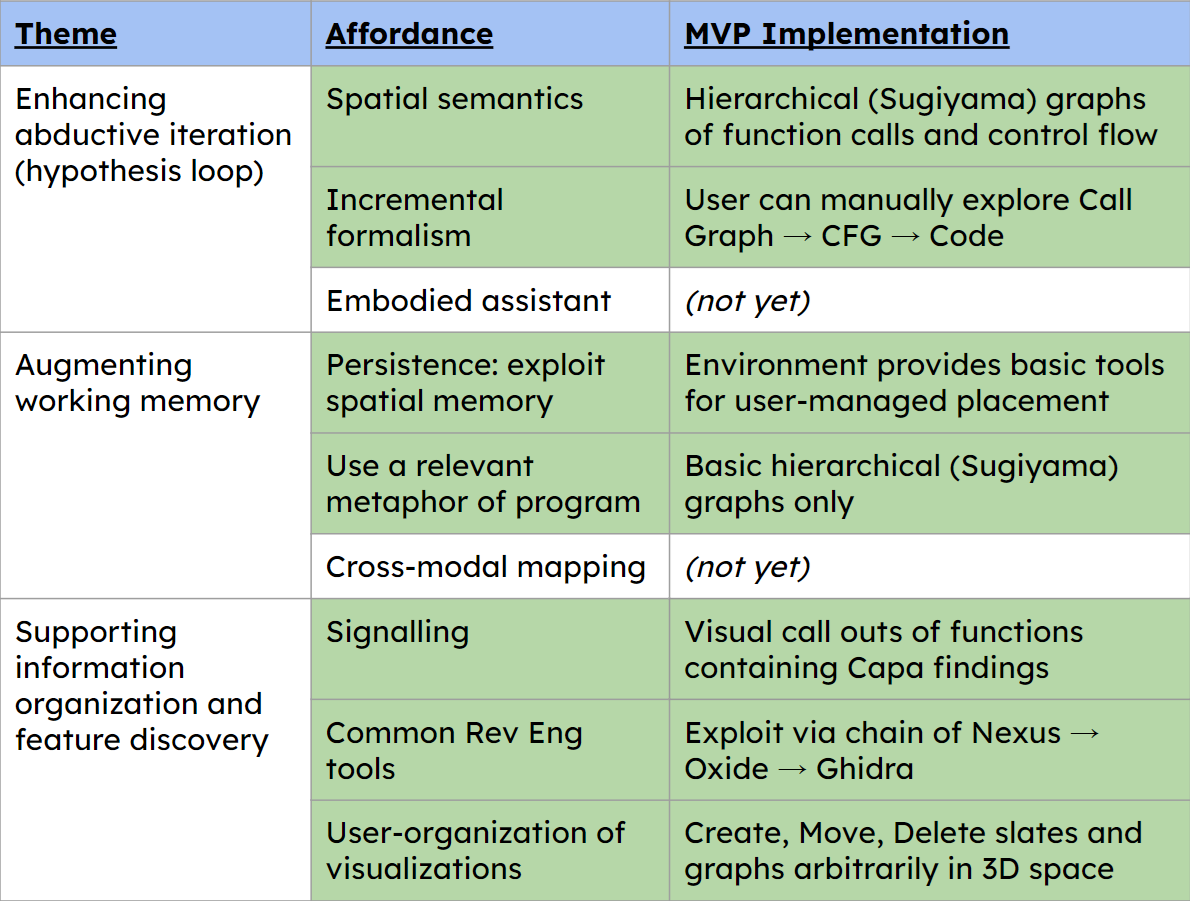}}
\caption{CogBRE MVP Feature Implementation.}
\label{fig:MVP}
\end{figure}

\section{System Design and Implementation}
\label{sec:implementation}

We designed and built a flexible platform, Cognitive Binary Reverse Engineering (CogBRE), in which to implement and evaluate relevant capabilities in an immersive environment. Based on our prioritized set of affordances, we defined a ``minimally viable product'' (MVP) feature set for CogBRE, shown in Figure~\ref{fig:MVP}. 

We employ two primary types of visualizations in the MVP, graphs and slates. Users can create, reposition, resize, and delete graphs and slates without limitation in the 3D space. 

\textbf{Graphs} visualize the elements and flows at two levels. \emph{Call graphs} display functions of a binary and the flows between them. \emph{Control Flow Graphs} (CFG) display basic blocks of a function and flows between them. The graphs can rendered via the hierarchical Sugiyama method~\cite{Healy2013} (as demonstrated in Figure~\ref{fig:screenshot}) or with a force-directed method. Users can select functions within these graphs for further analysis. Visual callouts draw the user's attention to functions with identifiable capabilities. 

\textbf{Slates} display scrollable outputs of tools brokered by Oxide and Nexus, such as header information, strings, function disassemblies, and function decompilations. We additionally have a singleton slate that allows user input and functions as a notepad.

\begin{figure}[htbp]
    \centerline{\includegraphics[width=0.5\textwidth]{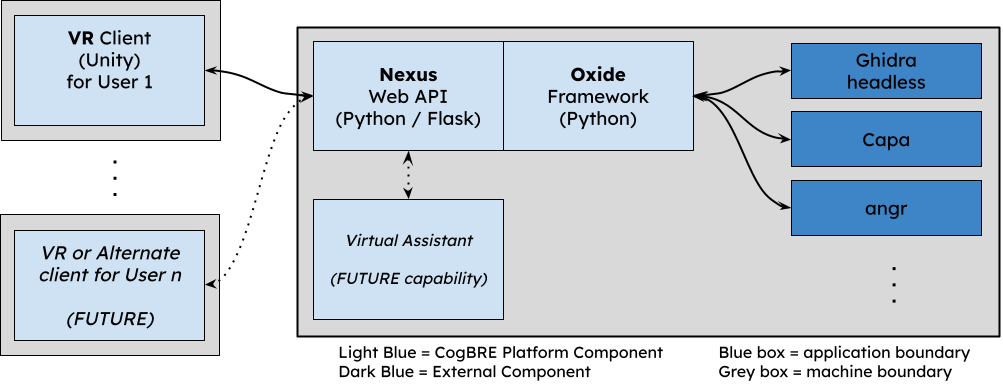}}
    \caption{CogBRE System Architecture.}
    \label{fig:architecture}
\end{figure}

We designed and selected a set of modular capabilities as shown in Figure~\ref{fig:architecture}. Oxide is a flexible tool for performing analysis of binary programs~\cite{mulder2014}; its design is modular, allowing the integration of multiple third-party tools along with custom analysis modules. The Nexus is the broker between the client systems and data sources through a basic RESTful web API. The VR client creates the immersive and interactive space for practitioners to explore data. Oxide provides this data by invoking common binary analysis tools on a selected binary program; while the choice of binary program is arbitrary, we are starting with x86 command line utilities. 

\section{Initial Feedback}
\label{sec:feedback}

We held sessions to solicit feedback from RE practitioners on the current system implementation. These sessions do not constitute a formal effectiveness study, which we plan to conduct later this year, but they give us valuable early feedback. This feedback was obtained from five different individuals associated with our research group (four graduate students and one undergraduate), having moderate or higher levels of experience with binary RE, and no to moderate experience with VR. We began our feedback sessions with a brief demonstration of the current capabilities of our program; users were then free to use the program as they saw fit, while we observed their experience through the lens of empathetic questions: \textit{What did they (users) do/say/think/feel?} This open-ended observation yielded these highlights: Users found the system easier to use in person than they expected based on prior videos and still frames of the system in action, and they needed little-to-no direction after donning the HMD and controllers. They remarked on the flexibility of the system compared to other reverse engineering tools discussed earlier, such as being able to analyze multiple binary files in the same environment. One individual who often feels nausea in VR did not in this case. Finally, users often mentioned illegibility of the smaller fonts; this is likely because the system was tuned for one particular HMD and, due to unfortunate circumstances, a different model with lower resolution was substituted for the feedback sessions.

Additionally, we employed a direct line of questioning to obtain explicit feedback. Summations of the feedback received are as follows:

\textbf{Would you use this tool as-is when trying to comprehend a binary program?} This question yielded mixed results; most (but not all) users said they would not use this program as is. Negative responders primarily cited poor text legibility (addressed above) and lack of functionality as to why they would not use this version of the program.  

\textbf{What is better in the VR environment than tools and processes you usually use? What is worse?} Users stated that obtaining information about the binary was significantly more intuitive than state-of-the-art tools discussed earlier.  Furthermore, users overwhelmingly cited that the vast spatial real-estate for management and comparison of different tasks supersedes the comparison capabilities present in the aforementioned tools.  In contrast, users lamented that the system offered no means to make annotations and notes in the environment.  Additionally, users stated that they wanted the program to match the functionality present in state of the art tools (e.g., function renaming, structure definition, library call identification, etc.).

\textbf{What are minor improvements you want to see in this tool?} Users predominantly stated that they wanted a means to integrate each abstraction into a larger unified graph, primarily through a ``find all instances'' function that takes in a selection of disassembly, and finds all instances of the assembly throughout each representation. Furthermore, some users wanted a means of condensing the information down into collapsible graphs that would progressively grow as they dived deeper into the binary. 

\textbf{With those minor improvements, would you use this tool when trying to comprehend a binary program?} Users unanimously stated that they would use this program with the aforementioned minor improvements, with the exception of one inconclusive answer.  Most users stated that they would apply this technology in the context of solving very basic reverse engineering tasks if these improvements were implemented.

\textbf{What are major improvements you want to see?} Users were widely varied in this inquiry. Most wanted to see advanced integraton with tools such as angr, and more robust comparisons between disassembly/decompilation across different functions.

\textbf{With those major improvements, would you use this tool when trying to comprehend a binary program?} Users predominantly stated that they would use CogBRE under the stipulation of improvement.

\section{User Study Design}
\label{sec:study}

We are planning a formal user study to discover quantifiable measures of effectiveness of using affordances in immersive VR in the process of performing binary RE tasks. The study protocol has been approved by our Institutional Review Board.  

The study will involve three conditions consisting of one traditional computer desktop environment and two immersive VR environments. The two VR environments will provide basic and advanced affordances. Participants, 18 for each condition, will execute one reverse engineering challenge in their assigned environment (this is a between-subjects design). In the traditional approach, they will use reverse engineering tools in a common desktop windowed environment controlled by keyboard and mouse. In the VR environments, they will use an approved head-mounted stereoscopic display and two hand controllers. 

Our hypotheses include: (1) Participants will have better task completion metrics with certain levels of affordances in VR; (2) Cognitive load will vary by level of affordances in VR; (3) Participants will rate certain affordances in VR (e.g., user organization, spatial memory, signalling) higher than others in terms of usefulness; and (4) Participants who have better performance in the immersive environments will have used more of the VR space.

\section{Conclusion and Future Work}
\label{sec:conclusion}

Based on research tying the cognitive challenges of binary RE with the cognitive benefits of VR, we designed and implemented the first iteration of an immersive platform on which we can evaluate the impact of affordances in VR on the binary RE process. Feedback on this initial implementation is encouraging and has given us valuable information for implementing the next iterations of our system. We are preparing to execute a formal human subjects research study to evaluate the relative impact of different affordances on solving binary RE tasks.

\bibliographystyle{IEEEtran}
\bibliography{VR}

\begin{thebibliography}{10}
\providecommand{\url}[1]{#1}
\csname url@samestyle\endcsname
\providecommand{\newblock}{\relax}
\providecommand{\bibinfo}[2]{#2}
\providecommand{\BIBentrySTDinterwordspacing}{\spaceskip=0pt\relax}
\providecommand{\BIBentryALTinterwordstretchfactor}{4}
\providecommand{\BIBentryALTinterwordspacing}{\spaceskip=\fontdimen2\font plus
\BIBentryALTinterwordstretchfactor\fontdimen3\font minus \fontdimen4\font\relax}
\providecommand{\BIBforeignlanguage}[2]{{%
\expandafter\ifx\csname l@#1\endcsname\relax
\typeout{** WARNING: IEEEtran.bst: No hyphenation pattern has been}%
\typeout{** loaded for the language `#1'. Using the pattern for}%
\typeout{** the default language instead.}%
\else
\language=\csname l@#1\endcsname
\fi
#2}}
\providecommand{\BIBdecl}{\relax}
\BIBdecl

\bibitem{Meng2016}
\BIBentryALTinterwordspacing
X.~Meng and B.~P. Miller, ``Binary code is not easy,'' in \emph{Proceedings of the 25th International Symposium on Software Testing and Analysis}, ser. ISSTA 2016.\hskip 1em plus 0.5em minus 0.4em\relax New York, NY, USA: Association for Computing Machinery, 2016, p. 24–35. [Online]. Available: \url{https://doi.org/10.1145/2931037.2931047}
\BIBentrySTDinterwordspacing

\bibitem{Wilson2002}
M.~Wilson, ``\BIBforeignlanguage{en}{Six views of embodied cognition},'' \emph{\BIBforeignlanguage{en}{Psychon Bull Rev}}, vol.~9, no.~4, pp. 625--636, Dec. 2002.

\bibitem{Hollnagel2005}
\BIBentryALTinterwordspacing
E.~Hollnagel and D.~Woods, \emph{Joint Cognitive Systems: Foundations of Cognitive Systems Engineering}.\hskip 1em plus 0.5em minus 0.4em\relax CRC Press, 2005. [Online]. Available: \url{https://books.google.vu/books?id=IwRHwOK2IzYC}
\BIBentrySTDinterwordspacing

\bibitem{Clark2008}
R.~Clark, D.~Feldon, J.~J.~G. Van~Merrienboer, K.~Yates, and S.~Early, ``Cognitive task analysis,'' in \emph{Handbook of Research on Educational Communications and Technology}, J.~M. Spector, M.~D. Merrill, J.~J.~G. van Merrienboer, and M.~P. Driscoll, Eds.\hskip 1em plus 0.5em minus 0.4em\relax New York: Routledge, 2008, ch.~43, pp. 577--593.

\bibitem{Brown2024arxiv}
\BIBentryALTinterwordspacing
D.~Brown, E.~Mulder, and S.~Mulder, ``Toward improving binary program comprehension via embodied immersion: A survey,'' 2024, arXiv:2404.17051. [Online]. Available: \url{https://arxiv.org/abs/2404.17051}
\BIBentrySTDinterwordspacing

\bibitem{Craik1943}
K.~J.~W. Craik, \emph{The Nature of Explanation}.\hskip 1em plus 0.5em minus 0.4em\relax Cambridge: Cambridge University Press, 1943.

\bibitem{Shneiderman1979}
\BIBentryALTinterwordspacing
B.~Shneiderman and R.~Mayer, ``Syntactic/semantic interactions in programmer behavior: A model and experimental results,'' \emph{International Journal of Computer {\&} Information Sciences}, vol.~8, no.~3, pp. 219--238, Jun 1979. [Online]. Available: \url{https://doi.org/10.1007/BF00977789}
\BIBentrySTDinterwordspacing

\bibitem{Bryant2012}
A.~Bryant, R.~Mills, G.~Peterson, and M.~Grimaila, ``Software reverse engineering as a sensemaking task,'' \emph{Journal of Information Assurance and Security}, vol.~6, pp. 483--494, 01 2012.

\bibitem{Nyre-Yu2022}
\BIBentryALTinterwordspacing
M.~Nyre{-}Yu, K.~Butler, and C.~Bolstad, ``A task analysis of static binary reverse engineering for security,'' in \emph{55th Hawaii International Conference on System Sciences, {HICSS} 2022, Virtual Event / Maui, Hawaii, USA, January 4-7, 2022}.\hskip 1em plus 0.5em minus 0.4em\relax ScholarSpace, 2022, pp. 1--10. [Online]. Available: \url{http://hdl.handle.net/10125/79608}
\BIBentrySTDinterwordspacing

\bibitem{Brooks1983}
\BIBentryALTinterwordspacing
R.~Brooks, ``Towards a theory of the comprehension of computer programs,'' \emph{International Journal of Man-Machine Studies}, vol.~18, no.~6, pp. 543--554, 1983. [Online]. Available: \url{https://www.sciencedirect.com/science/article/pii/S0020737383800315}
\BIBentrySTDinterwordspacing

\bibitem{Weigand2012}
K.~A. Weigand and R.~Hartung, ``Abduction's role in reverse engineering software,'' in \emph{2012 IEEE National Aerospace and Electronics Conference (NAECON)}, 2012, pp. 57--62.

\bibitem{Dudenhofer2019}
P.~P. Dudenhofer, ``Modeling and automating the cyber reverse engineering cognitive process,'' in \emph{23rd Colloquium for Information Systems Security Education}, ser. CISSE '19, 2019.

\bibitem{Votipka2020}
\BIBentryALTinterwordspacing
D.~Votipka, S.~Rabin, K.~Micinski, J.~S. Foster, and M.~L. Mazurek, ``An observational investigation of reverse {Engineers{\textquoteright}} processes,'' in \emph{29th USENIX Security Symposium (USENIX Security 20)}.\hskip 1em plus 0.5em minus 0.4em\relax USENIX Association, Aug. 2020, pp. 1875--1892. [Online]. Available: \url{https://www.usenix.org/conference/usenixsecurity20/presentation/votipka-observational}
\BIBentrySTDinterwordspacing

\bibitem{Sisco2017}
Z.~D. Sisco, P.~P. Dudenhofer, and A.~R. Bryant, ``Modeling information flow for an autonomous agent to support reverse engineering work,'' \emph{The Journal of Defense Modeling and Simulation}, vol.~14, no.~3, pp. 245--256, 2017.

\bibitem{Klein2007}
G.~Klein, J.~K. Phillips, E.~L. Rall, and D.~A. Peluso, ``A data-frame theory of sensemaking,'' in \emph{Expertise out of context: Proceedings of the Sixth International Conference on Naturalistic Decision Making}, R.~R. Hoffman, Ed.\hskip 1em plus 0.5em minus 0.4em\relax Lawrence Erlbaum Associates Publishers, 2007, pp. 113--155.

\bibitem{Zayour2000}
I.~Zayour and T.~C. Lethbridge, ``A cognitive and user centric based approach for reverse engineering tool design,'' in \emph{Proceedings of the 2000 Conference of the Centre for Advanced Studies on Collaborative Research}, ser. CASCON '00.\hskip 1em plus 0.5em minus 0.4em\relax IBM Press, 2000, p.~16.

\bibitem{Detienne2001}
F.~D\'{e}tienne and F.~Bott, \emph{Software Design---Cognitive Aspects}.\hskip 1em plus 0.5em minus 0.4em\relax Berlin, Heidelberg: Springer-Verlag, 2001.

\bibitem{Storey2005}
M.-A. Storey, ``Theories, methods and tools in program comprehension: past, present and future,'' in \emph{13th International Workshop on Program Comprehension (IWPC'05)}, 2005, pp. 181--191.

\bibitem{Mantovani2022}
A.~Mantovani, S.~Aonzo, Y.~Fratantonio, and D.~Balzarotti, ``{RE-Mind: a First Look Inside the Mind of a Reverse Engineer},'' in \emph{31st USENIX Security Symposium (USENIX Security 22)}.\hskip 1em plus 0.5em minus 0.4em\relax USENIX, 2022.

\bibitem{Scaife1996}
M.~Scaife and Y.~Rogers, ``External cognition: how do graphical representations work?'' \emph{International Journal of Human-Computer Studies}, vol.~45, no.~2, pp. 185--213, 1996.

\bibitem{Preece2019}
J.~Preece, Y.~Rogers, and H.~Sharp, \emph{Interaction Design: Beyond Human-Computer Interaction}, 5th~ed.\hskip 1em plus 0.5em minus 0.4em\relax Hoboken, NJ: Wiley, 2019.

\bibitem{Ale2022}
\BIBentryALTinterwordspacing
M.~Ale, M.~Sturdee, and E.~Rubegni, ``A systematic survey on embodied cognition: 11 years of research in child–computer interaction,'' \emph{International Journal of Child-Computer Interaction}, vol.~33, p. 100478, 2022. [Online]. Available: \url{https://www.sciencedirect.com/science/article/pii/S2212868922000174}
\BIBentrySTDinterwordspacing

\bibitem{Sweller2019}
J.~Sweller, J.~J.~G. Van~Merrienboer, and F.~Paas, ``Cognitive architecture and instructional design: 20 years later,'' \emph{Educational Psychology Review}, vol.~31, pp. 261--292, 06 2019.

\bibitem{Hollender2010}
N.~Hollender, C.~Hofmann, M.~Deneke, and B.~Schmitz, ``Integrating cognitive load theory and concepts of human–computer interaction,'' \emph{Computers in Human Behavior}, vol.~26, pp. 1278--1288, 11 2010.

\bibitem{Lisle2021}
L.~Lisle, K.~Davidson, E.~J. Gitre, C.~North, and D.~A. Bowman, ``Sensemaking strategies with immersive space to think,'' in \emph{2021 IEEE Virtual Reality and 3D User Interfaces (VR)}, 2021, pp. 529--537.

\bibitem{Helgesson2021}
D.~Helgesson and P.~Runeson, ``\BIBforeignlanguage{English}{Towards grounded theory perspectives of cognitive load in software engineering},'' in \emph{\BIBforeignlanguage{English}{Psychology of Programming Interest Group (PPIG)}}.\hskip 1em plus 0.5em minus 0.4em\relax Psychology of Programming Interest Group, Jun. 2021, psychology of Programming Interest Group Annual Workshop 2021 ; Conference date: 21-06-2021 Through 25-06-2021.

\bibitem{Andrews2010}
\BIBentryALTinterwordspacing
C.~Andrews, A.~Endert, and C.~North, ``Space to think: Large high-resolution displays for sensemaking,'' in \emph{Proceedings of the SIGCHI Conference on Human Factors in Computing Systems}, ser. CHI '10.\hskip 1em plus 0.5em minus 0.4em\relax New York, NY, USA: Association for Computing Machinery, 2010, p. 55–64. [Online]. Available: \url{https://doi.org/10.1145/1753326.1753336}
\BIBentrySTDinterwordspacing

\bibitem{deMelo2020}
\BIBentryALTinterwordspacing
C.~M. de~Melo, K.~Kim, N.~Norouzi, G.~Bruder, and G.~Welch, ``Reducing cognitive load and improving warfighter problem solving with intelligent virtual assistants,'' \emph{Frontiers in Psychology}, vol.~11, 2020. [Online]. Available: \url{https://www.frontiersin.org/articles/10.3389/fpsyg.2020.554706}
\BIBentrySTDinterwordspacing

\bibitem{Fittkau2015}
F.~Fittkau, A.~Krause, and W.~Hasselbring, ``Exploring software cities in virtual reality,'' \emph{2015 IEEE 3rd Working Conference on Software Visualization (VISSOFT)}, pp. 130--134, 2015.

\bibitem{Oberhauser2017}
\BIBentryALTinterwordspacing
R.~Oberhauser and C.~Lecon, ``Gamified virtual reality for program code structure comprehension,'' \emph{International Journal of Virtual Reality}, vol.~17, no.~2, p. 79–88, Jan. 2017. [Online]. Available: \url{https://ijvr.eu/article/view/2894}
\BIBentrySTDinterwordspacing

\bibitem{Capece2017}
N.~Capece, U.~Erra, S.~Romano, and G.~Scanniello, ``Visualising a software system as a city through virtual reality,'' in \emph{Augmented Reality, Virtual Reality, and Computer Graphics}, L.~T. De~Paolis, P.~Bourdot, and A.~Mongelli, Eds.\hskip 1em plus 0.5em minus 0.4em\relax Cham: Springer International Publishing, 2017, pp. 319--327.

\bibitem{Averbukh2019}
V.~Averbukh, N.~Averbukh, P.~Vasev, I.~Gvozdarev, G.~Levchuk, L.~Melkozerov, and I.~Mikhaylov, ``Metaphors for software visualization systems based on virtual reality,'' in \emph{Augmented Reality, Virtual Reality, and Computer Graphics}, L.~T. De~Paolis and P.~Bourdot, Eds.\hskip 1em plus 0.5em minus 0.4em\relax Cham: Springer International Publishing, 2019, pp. 60--70.

\bibitem{Romano2019}
\BIBentryALTinterwordspacing
S.~Romano, N.~Capece, U.~Erra, G.~Scanniello, and M.~Lanza, ``On the use of virtual reality in software visualization: The case of the city metaphor,'' \emph{Information and Software Technology}, vol. 114, pp. 92--106, 2019. [Online]. Available: \url{https://www.sciencedirect.com/science/article/pii/S0950584919301405}
\BIBentrySTDinterwordspacing

\bibitem{Hoff2022}
A.~Hoff, L.~Gerling, and C.~Seidl, ``Utilizing software architecture recovery to explore large-scale software systems in virtual reality,'' in \emph{2022 Working Conference on Software Visualization (VISSOFT)}, 2022, pp. 119--130.

\bibitem{Moloney2018}
J.~Moloney, B.~Spehar, A.~Globa, and R.~Wang, ``The affordance of virtual reality to enable the sensory representation of multi-dimensional data for immersive analytics: from experience to insight,'' \emph{Journal of Big Data}, vol.~5, no.~1, 2018.

\bibitem{Vessey1985}
I.~Vessey, ``Expertise in debugging computer programs: A process analysis,'' \emph{International Journal of Man-Machine Studies}, vol.~23, pp. 459--494, 1985.

\bibitem{Siegmund2014}
\BIBentryALTinterwordspacing
J.~Siegmund, C.~K\"{a}stner, S.~Apel, C.~Parnin, A.~Bethmann, T.~Leich, G.~Saake, and A.~Brechmann, ``Understanding understanding source code with functional magnetic resonance imaging,'' in \emph{Proceedings of the 36th International Conference on Software Engineering}, ser. ICSE 2014.\hskip 1em plus 0.5em minus 0.4em\relax New York, NY, USA: Association for Computing Machinery, 2014, p. 378–389. [Online]. Available: \url{https://doi.org/10.1145/2568225.2568252}
\BIBentrySTDinterwordspacing

\bibitem{Albus2021}
\BIBentryALTinterwordspacing
P.~Albus, A.~Vogt, and T.~Seufert, ``Signaling in virtual reality influences learning outcome and cognitive load,'' \emph{Computers \& Education}, vol. 166, p. 104154, 2021. [Online]. Available: \url{https://www.sciencedirect.com/science/article/pii/S0360131521000312}
\BIBentrySTDinterwordspacing

\bibitem{Mayer2005}
R.~E. Mayer, \emph{Cognitive Theory of Multimedia Learning}, ser. Cambridge Handbooks in Psychology.\hskip 1em plus 0.5em minus 0.4em\relax Cambridge University Press, 2005, p. 31–48.

\bibitem{Batch2020}
A.~Batch, A.~Cunningham, M.~Cordeil, N.~Elmqvist, T.~Dwyer, B.~H. Thomas, and K.~Marriott, ``There is no spoon: Evaluating performance, space use, and presence with expert domain users in immersive analytics,'' \emph{IEEE Transactions on Visualization and Computer Graphics}, vol.~26, no.~1, pp. 536--546, 2020.

\bibitem{Healy2013}
P.~Healy and N.~Nikolov, \emph{Hierarchical Drawing Algorithms}.\hskip 1em plus 0.5em minus 0.4em\relax CRC Press, 08 2013, pp. 409--454.

\bibitem{mulder2014}
\BIBentryALTinterwordspacing
S.~A. Mulder, ``Cross-domain situational awareness in computing networks,'' \emph{Sandia National Laboratories (SAND Report)}, 11 2014. [Online]. Available: \url{https://www.osti.gov/biblio/1494613}
\BIBentrySTDinterwordspacing

\end{thebibliography}

\end{document}